\documentclass[%
 preprint,
 superscriptaddress,
 amsmath,amssymb,
 aps,
 pra,
floatfix,
]{revtex4-2}

\usepackage[english]{babel}

\usepackage{amsmath}
\usepackage{graphicx}
\usepackage[colorlinks=true, allcolors=blue]{hyperref}
\usepackage{xcolor}
\usepackage{booktabs}
\usepackage[normalem]{ulem}
\usepackage{array}[=2016-10-06]

\usepackage{siunitx}

\begin{document}
\title{Origins of microwave losses in superconducting circuits made with silicon-on-insulator substrates}

\author{Simon Messelot}
\author{Nicolas Aparicio}
\author{Kazi Rafsanjani Amin}
\author{Eric Eyraud}
\affiliation{Univ. Grenoble Alpes, CNRS, Grenoble INP, Institut N\'eel, 38000 Grenoble, France}
\author{Bruno Fain}
\author{Mikaël Cassé}
\author{Guillaume Jourdan}
\author{Fabrice Nemouchi}
\author{Sébastien Hentz}
\affiliation{Univ. Grenoble Alpes, CEA, LETI, 38000 Grenoble, France}
\author{Frédéric Gustavo}
\author{François Lefloch}
\affiliation{Univ. Grenoble Alpes, CEA, Grenoble INP, IRIG, PHELIQS, 38000 Grenoble, France
}
\author{Nicolas Roch}
\author{Jérémie J. Viennot}
\author{Julien Renard}\email{julien.renard@neel.cnrs.fr}
\affiliation{Univ. Grenoble Alpes, CNRS, Grenoble INP, Institut N\'eel, 38000 Grenoble, France}


\begin{abstract}

Silicon-on-insulator technology is widely used to fabricate silicon based devices, from advanced transistors to photonic circuits or nanomechanical systems. Integrating low loss superconducting quantum circuits with silicon-on-insulator substrates enables to couple the advantages offered by the mature silicon technology to the exquisite sensitivity of superconducting circuits. The natural approach, inherited from research in superconducting microwave devices, is to use a substrate made with highly resistive silicon, known for its low level of microwave losses. In this work, using superconducting microwave resonators, we show that counterintuitively, standard resistivity silicon-on-insulator substrates perform better than high resistivity silicon-on-insulator substrates at cryogenic temperatures. In the latter case, the presence of a parasitic sheet conduction at the interface between bulk silicon and silicon oxide acts as the dominant loss mechanism. 
This parasitic sheet can be suppressed using substrates with intentionally induced traps. In such substrates, losses are ultimately limited by the dielectric losses of the silicon oxide layer. These substrates offer interesting perspectives for the development of superconducting nanoelectromechanical systems. First, the release, i.e. the removal of the silicon oxide, could be limited to the moving parts, thereby maintaining the mechanical integrity of the rest of the device. Additionally, such structure would enhance heat evacuation into the bulk of the substrate which is an issue in current devices such as microwave-to-optics converters. 
\end{abstract}

\maketitle

\section{Introduction}

Silicon-on-insulator (SOI) substrates have been used recently in different fields of quantum technologies: photonics \cite{Silverstone2015}, electromechanics \cite{dieterle2016superconducting}, optical-to-microwave transducers \cite{zhao2025quantum, arnold2020converting}, silicon based spin quantum bits (qubits) \cite{Maurand2016,CardosoPaz2024} or superconducting qubits \cite{keller2017transmon}. Fabrication on such substrates can benefit from the mature silicon processing technologies, while offering advantages for nanomechanical systems, in which the buried silicon oxide layer can naturally serve as a sacrificial layer during the release of the structure \cite{cleland2025nanoscalemechanicalstructuresfabricated}.

Superconducting microwave resonators can exhibit high quality factors exceeding $10^6$ at the single photon level, making them very sensitive tools for quantum technologies and also to study microwave dissipation channels in various platforms \cite{sage2011study}. These include for instance different superconducting materials such as aluminum, niobium or tantalum \cite{lozano2024, van2025}, substrates \cite{Checchin2022,Read2024,Shen2024} or interfaces \cite{bruno2015,woods2019}. 

Detailed investigations of microwave losses in resonators on bulk substrates (silicon or sapphire) have shown that the dominant loss mechanisms are often linked to the presence of two-level systems (TLS) at interfaces \cite{sage2011study, mcrae2020materials}. When using  high kinetic-inductance superconducting materials, such as ultra-thin TiN or granular aluminum, an additional mechanism, due to the presence of localized quasi-particles, can contribute significantly \cite{Grunhaupt2018,amin2022loss}. 
Advanced devices including superconducting resonators, such as microwave-to-optics transducers operating at cryogenic temperatures, have already been demonstrated on SOI substrates. 
It was originally shown that the presence of the buried oxide combined with highly resistive silicon substrates appeared detrimental to the superconducting resonator quality factor \cite{dieterle2016superconducting}, well beyond what can be expected from known sources of loss in silicon oxide (SiO$_2$). The choice was made to remove the buried oxide in large portions of the device to suppress this effect \cite{dieterle2016superconducting,keller2017transmon}. This release of the top Si layer results nevertheless in more fragile structures and decreases the thermal coupling to the bulk substrate. However no dedicated study of microwave loss mechanisms specific to SOI substrates was reported at cryogenic temperature.\\

Properties of SOI substrates have been extensively studied to develop high frequency electronics based on CMOS transistors \cite{raskin2022fully}, down to 100 mK \cite{galy2018cryogenic}. In terms of microwave performances at room temperatures, limitations have been attributed to the existence of a parasitic surface conduction (PSC) channel at the (bottom) silicon-SiO$_2$ interface \cite{Wu99}, resulting in ohmic losses. This conduction layer stems from positive charges located in the buried oxide layer, typically in the range of 10$^{10}$ to 10$^{12}$ cm$^{-2}$, attracting mirror charge carriers at the interface with silicon. The properties of the PSC is highly dependent on the substrate doping type. In p-type substrates, the presence of positive charges in the buried oxide layer leads to the creation of an inversion layer close to the interface and a related depletion region farther away from the interface. At room temperature, local carrier densities exceeding 10$^{17}$ cm$^{-3}$ can exist \cite{Berlingard2023}, resulting in a local resistivity below $\SI{1}{\ohm \cdot \centi\meter}$ close to the interface, i.e. smaller than the bulk resistivity for both standard and high-resistivity SOI substrates. Nevertheless at room temperature high-resistivity SOI substrates still perform better in radiofrequency applications than the standard ones due to the reduced bulk substrate losses and crosstalk. \\

The situation changes when lowering the temperature. At cryogenic temperature, losses due to the presence of bulk free carriers induced by dopants are strongly suppressed, even for standard resistivity SOI substrates. Dopant freeze-out indeed exponentially increases the bulk resistivity below 50~K. The PSC, whose exact properties are strongly influenced by the substrate doping level,  becomes the dominant loss channel \cite{Berlingard2023,berlingard2024caracterisation}. It results in a performance crossover when lowering the temperature. At 7~K standard SOI substrates end up exhibiting better performances regarding microwave losses compared to high-resistivity SOI \cite{Berlingard2023,berlingard2024caracterisation}.
However, suppression of the PSC can be achieved thanks to the inclusion of a thin layer of polycrystalline silicon beneath the buried oxide, that acts as traps suppressing the PSC mobility. Significant decrease of microwave losses have been demonstrated with this technique in so called trap-rich SOI substrates. They are today the highest performance standard for microwave devices at room temperature  \cite{neve2012small,Raskin2018}. Additionally, a recent study indicates that the presence of a trap rich layer can also effectively neutralize the PSC in lithium niobate on insulator substrates \cite{Shen2024}.

In this article, we present a comprehensive and comparative study of the performances of SOI substrates using superconducting circuits at cryogenic temperatures. We fabricate superconducting microwave resonators made of TiN and study the influence photon number and temperature on the internal quality factor, which quantitatively characterizes substrate induced losses. 
We show that high-resistivity SOI substrates present large losses, due to the presence of the PSC layer, limiting internal quality factors to less than $10^3$. Standard resistivity SOI substrates perform better but they are limited by the presence of losses due to the residual doping in the Si substrate. Trap-rich SOI substrates, for which these two mechanisms are effectively suppressed, significantly outperform both. They exhibit internal quality factors above $2.10^4$ in the single photon limit, only limited by losses in the SiO$_2$ layer.

\section{Device description and basic characterization}
\begin{figure*}[t]
\centering
\includegraphics[width=11.8cm]{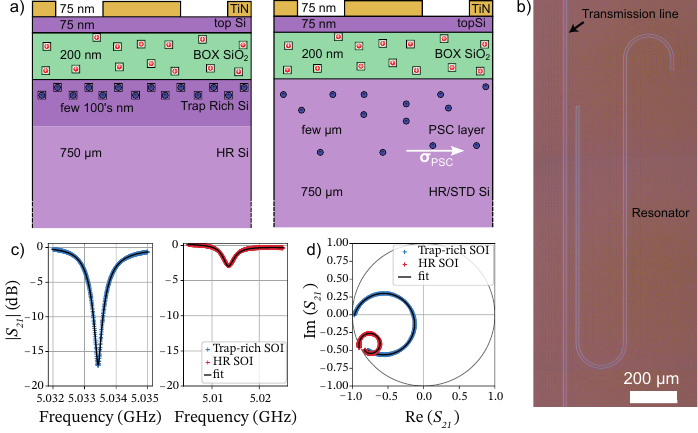}
\caption{\label{fig:Presentation} \textbf{Properties of the studied devices} \textbf{a)} Schematic stacks of the studied devices for SOI trap-rich (left) and high-resistivity SOI or standard SOI (right) substrates. Carriers in square boxes denote trapped charges while carriers in circle are free carriers representing the parasitic surface conduction layer (PSC) that can exist in both high-resistivity (HR) and standard (STD) SOI substrates. This layer can have a significant conductivity $\sigma_{PSC}$.  \textbf{b)} Optical micrograph of the hanger type resonator. The TiN thin film is in brown. The bright area are the narrow gaps defining the CPW transmission line (left part) and resonator (meander). The resonator width is 10~$\mu$m and the gap is 2~$\mu$m.  \textbf{c)} Typical resonance spectra measured at high power for SOI trap-rich (left), and high-resistivity SOI (right), with fitted curves in black. \textbf{d)} Resonance curves in the Re$(S_{21})$/Im$(S_{21})$ plane with their respective fits.}
\end{figure*}

The structure of a SOI substrate consists of an $\mathrm{SiO_2}$ insulating layer (the buried oxied, BOX) sandwiched between a thin layer of silicon on top (Si-top) and a bulk, crystalline silicon substrate below (referred to as Si handle). We study the microwave loss properties of three different SOI substrates in addition to a reference highly resistive bulk silicon substrate (HR Bulk). A first SOI substrate contains a highly resistive Si handle (high-resistivity SOI). The second SOI substrate type is made with a standard resistivity Si handle (standard SOI). Finally, the third SOI substrate consists of a highly resistive Si handle and an additional thin layer of polycrystalline silicon at the bottom SiO$_2$ interface (Trap-rich SOI). The different substrate stacks are represented in Fig. \ref{fig:Presentation}.a) and their characteristic details  (layer thicknesses and resistivity) are summarized in Table \ref{tab:substrate}.

\begin{table}[ht]
\begin{center}
\begin{tabular}{|p{6cm}|p{2cm}|p{2cm}|p{3cm}|p{2cm}|}

\hline
Substrate description & Acronym & $t_{Si,top}$ (nm) & $t_{SiO_2,BOX}$ (nm) & $\rho_{Bulk}$ \\
\hline
\hline
High resistivity silicon  & HR Bulk &$\O$ & $\O$ & $\SI{4000 }{\ohm \cdot \centi\meter}$ \\
\hline
Silicon on insulator with high resistivity bulk& HR SOI &75 & 200 & $\SI{2000}{\ohm \cdot \centi\meter}$ \\
\hline
Silicon on insulator with standard resistivity bulk\ & Std SOI &88 & 145 & $\SI{10}{\ohm \cdot \centi\meter}$ \\
\hline
Silicon on insulator with a trap rich layer  and high resistivity bulk& Trap-rich SOI &75 & 200 & $\SI{5800}{\ohm \cdot \centi\meter}$ \\
\hline

\end{tabular}
\end{center}
\caption{Details of the different substrates studied. $t$ represents the thickness of the top silicon layer ($t_{Si,top}$) and SiO$_2$ buried oxide ($t_{SiO_2,BOX}$). $\rho_{Bulk}$ is the room temperature resistivity of the silicon substrate (HR Bulk) or Si handle (SOI substrates). For all SOI substrates, the top silicon layer room temperature resistivity is $\rho_{Si,top}\approx$10~$\Omega$.cm. The doping is p-type for all substrates.}
\label{tab:substrate}
\end{table}

All substrates underwent an identical process to fabricate $\lambda / 4$ superconducting TiN resonators in a coplanar waveguide (CPW) geometry as well as test structures for direct current (DC) chararectization. The resonators are capacitively coupled to a microwave transmission line, as shown in Fig.~\ref{fig:Presentation}.b. TiN was chosen as a superconducting material because of its compatibility with 300~mm semiconductor facilities. A 75~nm-thick TiN layer was deposited on the full wafer by physical vapor deposition (PVD), over a thin 5 nm Ti layer used to improve TiN quality.
The Ti/TiN layer was then patterned to form the resonator and DC test structures using deep UV lithography and dry etching. Compared to our previous work on TiN which used much thinner layers \cite{amin2022loss}, the thickness is large here to suppress localized quasiparticle induced losses which dominate in high kinetic inductance thin film devices \cite{grunhaupt2018loss}. 

First, all wafers were characterized at 300~K, with a mapping of the TiN resistivity over the 300~mm wafer, using dedicated van der Pauw structures. All the wafers exhibited an uniformity better than 4{\%}  across the wafer. The critical temperature of the TiN layer deposited using the same process has been evaluated in a full-sheet experiment, consisting of a non-patterned TiN layer. We characterized the 75~nm thick TiN layer temperature-dependent resistivity and measured a critical temperature $T_c = 3.7$ K and normal state resistance $R_N = \SI{26.0}{\ohm} / \square$. From these measurements, we estimate the kinetic inductance $L_k = \SI{9.7}{\pico\henry}/ \square$ using $L_K =\hbar R_N / \pi \Delta_0$ \cite{shearrow2018atomic,samkharadze2016high} and the BCS formula $\Delta_0 = 1.76 k_B T_C$ to estimate the superconducting gap.\\

We perform the characterization of the superconducting resonators in a dilution refrigerator, with a base temperature of \SI{25}{\milli\kelvin}. Measurements are performed at this base temperature, unless stated otherwise. Using a vector network analyzer, we probe the $S_{21}$ scattering parameter of a superconducting co-planar waveguide (CPW) transmission line capacitively coupled to the resonator, and measure the fundamental resonant mode. Typical resonance shapes are represented in Fig.~\ref{fig:Presentation}.c and d. From the measured resonance frequency $f_r \simeq 5.02$ GHz and the resonator geometry, we estimate $L_k$ with electromagnetic simulations, performed with the Sonnet software. The experimental resonance frequency is obtained for $L_k = \SI{9.95}{\pico\henry} / \square$, which is very close, within 2\%, to the value obtained from DC resistivity measurements.  Simulation of the same resonator geometry considering only the geometric inductance $L_0$ (with $L_k = \SI{0}{\pico\henry} / \square$) yields $f_0 = \SI{10.88}{\giga \hertz}$. We deduce then the kinetic inductance fraction $\alpha = \frac{L_k}{L_k + L_0} = 1 - \left( \frac{f_r}{f_0} \right)^2 = 78.7\% $ an important quantity for the analysis of the effect of thermal quasiparticles, which will be presented later.\\

Losses in the device are characterized by the internal quality factor $Q_i$ {of the resonator, that we extract using standard fitting procedures in the complex plane of the $S_{21}$ parameter (Fig.~\ref{fig:Presentation}.d). The model used for fitting is \cite{probst2015efficient}:

\begin{equation}
    S_{21}(f) = A e^{j \phi_{offset}} \left[ 1 - \frac{ (Q/Q_c) e^{j\phi} }{ 1 + 2jQ(f/f_r-1)} \right]
\end{equation}

where $f_r$ is the resonance frequency, $Q_c$ is the coupling quality factor to the transmission line, $Q$ is the total quality factor with $Q^{-1} = Q_c^{-1} + Q_i^{-1}$. $A$ and $\phi_{offset}$ describe baseline amplitude and phase. $e^{j\phi}$ is an additional parameter describing impedance mismatches in the transmission line, originating from the kinetic inductance of the TiN film or setup imperfections.
We set in a first design $Q_c = 5900$, and in the following, we measure values ranging between 5100 and 5800 depending on sample details.

In Fig. \ref{fig:Presentation}.c), we observe that the resonance dips can have very different contrast depending on the substrates. This is an indication of large differences in  $Q_i$. In the left plot for the trap-rich SOI substrate sample, the minimal value of  $|S_{21}|$  reaches -17.1 dB, indicating $Q_i \gg Q_c$. This configuration leads to a large uncertainty in the determination of the internal quality factor $Q_i$. To tackle this issue, we proceeded to additional measurements on similar resonators with $Q_c$ increased by design to 53 000 ($Q_{c+}$), for both trap-rich SOI and HR bulk substrates, that are reported in the following.

\section{Microwave power dependence}
We first discuss the power dependent losses in the four types of substrates. The relevant metric for the microwave radiation intensity is the average number of photons circulating in the resonator $<n_{photon}>$, estimated from the input power on the sample $P_{in}$:

\begin{equation}
    <n_{photon}> = \frac{ P_{in} }{ \hbar \pi^2 f_r^2 } \frac{Q^2}{Q_c} 
\end{equation}\\

\begin{figure*}[t]
\centering
\includegraphics[width=\textwidth]{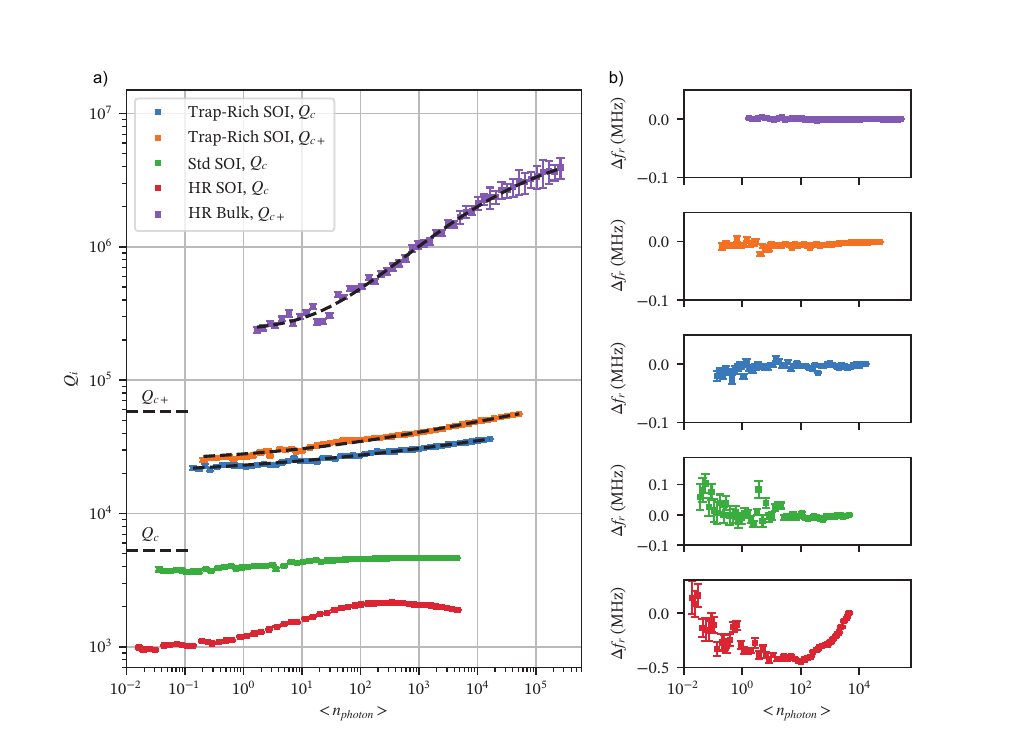}
\caption{\label{fig:Power} \textbf{Power dependence of internal quality factors Q$_i$ and resonance frequency shifts $\Delta f_r$.} Data taken at 25~mK. \textbf{a)} $Q_i$ for the different substrates. Dashed lines are fits using a TLS based model described in the text for HR Bulk and trap-rich SOI substrates. $Q_c=5300$ and $Q_{c+}=58000$ are the two different external quality factors used in this study. \textbf{b)} Shift of the resonance frequency with photon number: $\Delta f_r = f_r - f_{r,ref}$ (Reference is taken at high power).}
\end{figure*}

The power dependence of $Q_i$ is presented in Fig. \ref{fig:Power}.
In Fig. \ref{fig:Power}.a), we first observe that globally, substrates rank from lower to higher quality factors as high-resistivity SOI, standard SOI, trap-rich SOI and finally HR bulk. Trap-rich SOI is the best substrate within SOI substrates, high-resistivity SOI performs worse than standard resistivity SOI, which suggests a signature of PSC associated losses. The HR bulk substrate performs significantly better since it is not limited by  additional mechanisms present in the SOI stack. $Q_i$ data for trap-rich SOI substrates obtained using $Q_c$ values of $\sim$ 58 000 and 5 300 yield similar values, yet with a lower error bar for $Q_c$ $\sim$ 58 000,  supporting our analysis.\\

For all samples, we observe saturable absorption, i.e. an increase of the internal quality factor $Q_i$ with increasing photon population in the resonator. Nevertheless, the magnitude of the effect is strongly substrate dependent. This may be explained by the presence of localized two-level systems (TLS) responsible for dissipation in the single photon limit. Evolution of $Q_i$ due to the saturation of TLS with increasing photon number or temperature can be fitted using \cite{martinis2005,Gao2008,Crowley2023}:

\begin{equation}
    Q_{i,tot}^{-1} = Q_{i,sat}^{-1} \times \frac{\tanh(\frac{\hbar \omega}{2 k_B T})}{\sqrt{1+\left(\frac{<n>}{n_c}\right)^{\beta}\tanh(\frac{\hbar \omega}{2 k_B T})}}  +   Q_{i,other}^{-1}
    \label{eq:TLS}
\end{equation}

where $Q_{i,sat}^{-1}$ is the quality factor associated to saturable dissipation, $n_c$ is the photon occupancy threshold for saturation effects, $Q_{i,other}^{-1}$ describes non-saturable dissipation and $\beta$ is an empirical parameter, which is equal to 1 for dielectric loss induced by TLS saturation in an homogeneous electric field \cite{VonSchickfus1977,martinis2005} and 0.8 in a coplanar waveguide geometry for losses coming from the metal surface \cite{wang2009}. In the low temperature, low photon number limit, the intrinsic quality factor reads $Q_{i,single}^{-1} = Q_{i,sat}^{-1} + Q_{i,other}^{-1}$. Fitting parameters are reported in Table \ref{tab:fit_power} for trap-rich SOI and HR bulk substrates, whose power dependence is well captured by this model.\\

\begin{table}[ht]
\centering
\begin{tabular}{|l|l|l|l|l|}
\hline
Substrate & $Q_{i,sat} $ & $< n_c >$ & $\beta$ & $Q_{i,other}$ \\
\hline
Trap-rich SOI & $19~000\pm4~000 $ & $48\pm24$ & $0.17\pm0.04$ & $> 100~000$  \\
Trap-rich SOI, Qc+ & $28~000\pm1~000 $ & $37\pm8$ & $0.27\pm0.01$ & 190~000$\pm$30~000 \\
HR bulk, Qc+ & $245~000\pm9~000$ & $20\pm3$ & $0.81\pm0.03$ & $5~800~000\pm900~000$ \\
\hline
\end{tabular}
\caption{Fitting parameters obtained using the TLS model of Eq.~\ref{eq:TLS} for the power dependent measurement of the quality factors.}

\label{tab:fit_power}
\end{table}

$Q_i$ in the HR bulk substrate reaches about 250 000 at the single photon level and it approaches a plateau with $Q_{i,other} \approx 6 \times 10^6$ at large photon number, $<n_{photon}> ~ \sim10^5$  .
The increase of $Q_i$ with photon number by more than one order of magnitude indicates that the saturable absorption process is the dominant loss mechanism.
The slope of increase is characterized by the $\beta$ parameter, with $\beta = 0.81$ being close to the value expected for saturable absorption due to TLS localized at interfaces in a CPW geometry with an homogeneous substrate \cite{wang2009}.

$Q_i$ in trap-rich SOI substrates is one order of magnitude lower than in HR bulk substrates in the single photon limit with $Q_{i,single} \approx$ 25 000 for the sample with the highest $Q_c$.
The large difference compared to the HR bulk substrate reference indicates that losses in trap-rich SOI substrates are dominated by specific mechanisms not present in HR bulk substrates. $Q_i$ in trap-rich SOI substrates exhibits a noticeable increase with power, but the slope is here much smaller than in the HR bulk substrate, with $\beta  \sim 0.3$. This could have several origins. First, SOI substrates have a non-homogeneous dielectric medium. Different medium and interfaces will experience different electric fields which is expected to influence $\beta$. Second a reduced $\beta$ could mean that part of the absorption mechanisms at play are non-saturable, or that the TLS bath is so large that it requires a very large number of photon to be saturated. Indeed, even for $<n_{photon}> \sim 10^5$, $Q_i$ does not reach any clear plateau despite increasing up to 55 000. For this reason, it is difficult to precisely quantify the ultimately reachable value $Q_{i,other}$ in these trap-rich SOI substrates.

We also obtain different critical photon numbers $n_c$ for trap-rich SOI substrates compared to HR bulk substrates, albeit with larger uncertainties. $n_c$ is expected to depend on several parameters: the electric dipole of the TLS, their relaxation and coherence time, as well as the value of the electric field at their locations \cite{Gao2008}. At this point, we cannot conclude on the main factor causing the differences in $n_c$ for the two types of substrates.\\

Turning now to standard SOI and high-resistivity SOI substrates, we found that the observed power dependence cannot be fitted using the TLS model of Eq. \ref{eq:TLS}. $Q_i$ in standard SOI substrates only weakly increases with  power (by 20\%), which suggests a non-saturable dominant loss mechanism by opposition to localized saturable defects such as TLS.

For the high-resistivity SOI substrate, we observe an increase by a factor 2 of $Q_i$, followed by a small but noticeable decrease for $<n_{photon}>$ above $3\times10^{2}$. Such variations cannot be described by TLS models. We note in Fig.~\ref{fig:Power}.b) that only high-resistivity SOI exhibits a significant power-induced frequency shift of up to 0.4 MHz, which could be associated to parasitic sheet capacitance linked to PSC. These two observations taken together points towards a modulation of the PSC properties by the microwave power as the dominant mechanism for the high-resistivity SOI substrate. Nevertheless the mechanism remains undetermined at this stage, since the PSC is not expected to exhibit saturable absorption per se.

\section{Temperature dependence}

We proceed to further investigation of the loss mechanisms by studying the temperature dependence of $Q_i$ for all substrates.
$Q_i$ measurements versus temperature are reported in Fig. \ref{fig:Temp}, where we included both high microwave power and single-photon regime data.\\

\begin{figure*}[t]
\centering
\includegraphics[width=\textwidth]{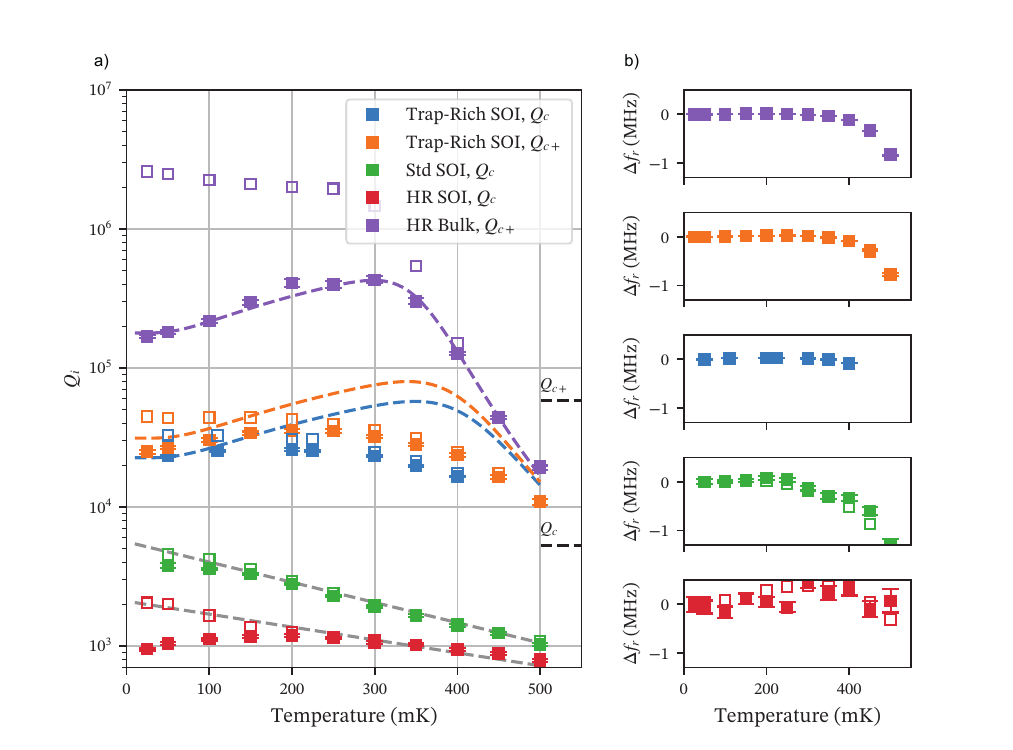}
\caption{\label{fig:Temp} \textbf{Temperature dependence of internal quality factor and resonance frequency.} Hollow marker data are taken at high microwave power ($<n_{photon}> \sim 10^{4}$), full marker data are taken at low microwave power ($<n_{photon}> \sim 10^{-1}$). \textbf{a)} $Q_i$ for the different substrates. Colored dashed lines use a model of TLS (Eq. \ref{eq:TLS}) and QP (Eq. \ref{eq:QP}) losses versus temperature at low power for HR bulk substrates (fitted parameters) and trap-rich SOI substrates (fixed parameters) substrates. Gray dashed lines are a fit by an empirical $\exp(-T/T_X)$ law for standard SOI and high-resistivity SOI substrates at high power.  For HR bulk substrate data at high power, $Q_i$ at low $T$ is 50 times higher than $Q_c$, so that $Q_i$ estimates only provide an order of magnitude.  \textbf{b)}  Shift of the resonance frequency with temperature $\Delta f_r = f_r - f_{r,ref}$ (Reference is taken at 25 mK).}
\end{figure*}

We observe that overall, the differences between low and large photon numbers is reduced when increasing the temperature. This comes from two effects. First temperature and photon occupancy play similar roles regarding saturable absorption (see Eq.~\ref{eq:TLS} for TLS for example). Second, at higher temperature, microwave losses are dominated by other, thermally activated, mechanisms.  
Focusing on the single-photon data, the reference HR bulk substrate shows a large increase of $Q_i$ with temperature up to $T = \SI{300}{\milli\kelvin}$ with $Q_i = 430~000$ followed by a dramatic drop above this temperature. The increase of $Q_i$  results from the thermal excitation of the TLS, thereby reducing their effect on microwave losses (see Eq.\ref{eq:TLS}) \cite{VonSchickfus1977}. On the other hand,  the  decrease of $Q_i$ observed above \SI{300}{\milli\kelvin} can be attributed to thermal quasiparticles, responsible for microwave losses. The contribution of thermal quasiparticles to the quality factor is given, in the thin film limit, by \cite{Gao2008b,Zmuidzinas12,Alexander2025}:

\begin{equation}
    Q_{i,QP}^{-1} = \frac{4 \alpha}{\pi} \frac{\Delta(T)}{\Delta_0} \sinh(\frac{\hbar \omega}{2 k_B T}) K_0(\frac{\hbar \omega}{2 k_B T}) \exp(- \frac{\Delta (T)}{k_B T})
    \label{eq:QP}
\end{equation}

where $\alpha$ is the kinetic inductance fraction of the resonator, $\Delta(T)$ ($\Delta_0$) the temperature dependent superconducting gap (zero temperature superconducting gap) and $K_0$ the zeroth-order modified Bessel function of the second kind. We fit the low power data for the HR bulk substrate with a model including TLS and thermally activated quasi-particles (QPs), leaving $\Delta_0$ as a free parameter and using $\alpha$ estimated from electromagnetic simulations. We found a good agreement with experimental data with $\Delta_0 / h=$ 95 GHz, which is comparable to the gap estimated from the critical temperature using the BCS formula ($\Delta_0 / h = \SI{132}{\giga\hertz}$).\\

In trap-rich SOI substrates, $Q_i$ exhibits a small but significant bell shape versus temperature in the single photon regime, meaning that we observe an increase of $Q_i$ with increasing temperature before seeing a significant decrease above \SI{250}{\milli\kelvin}. Similarly to power dependent measurements, the increase of the quality factor with temperature indicates saturation of TLS through thermal excitation. However, contrary to the HR bulk substrate, we cannot fit the decrease of $Q_i$ using the model of thermal QPs with reasonable parameters. Indeed, the losses due to thermal QPs are associated to the TiN superconducting resonator, which is identical in HR bulk and trap-rich SOI substrates. QPs associated losses are thus expected to be quantitatively similar for both substrates. We observe on the contrary a significant decrease of $Q_i$ for trap-rich SOI substrates at 300 - 350 mK despite $Q_i$ due to thermal QPs being one order of magnitude higher, i.e. negligible compared to the dominant dissipation mechanism.

As an illustration, we show in Fig. \ref{fig:Temp} in orange and blue dashed lines the expected $Q_i$ for trap-rich SOI substrates assuming contributions from both a TLS bath and thermal QPs described by Eq. \ref{eq:TLS} and \ref{eq:QP} respectively. We use $Q_{i,sat}$ extracted from the fit on trap-rich SOI substrates power dependence data and $\Delta_0$ from the fit on HR bulk substrates temperature dependence data. The initial increase is well accounted for with this model but it fails to capture accurately the decrease with temperature above 250 mK. Consequently, another temperature dependent mechanism is at play in trap-rich SOI substrates. A potential candidate is the layer of polycristalline silicon specific to the trap-rich SOI substrates. Losses in that layer, designed to trap free carriers, could be thermally activated. Also the limited resistivity of the top silicon layer ($\approx$10~$\Omega$.cm) could be an additional source of thermally activated losses.

The temperature evolutions of $Q_i$ for standard SOI and high-resistivity SOI substrates are very different from what we measured on other substrates. The temperature dependence of $Q_i$ for the standard SOI substrate follows an empirical exponential law as shown by the fit $\propto e^{-T/T_X}$ on high power data, with $T_X\approx$ 300~mK for standard SOI. This dependence is not compatible with usual model for losses, and its potential origin is discussed further below. The significant bell shape observed in the high-resistivity SOI substrate at low power is in agreement with the saturable absorption observed in Fig. \ref{fig:Power}. In the high power data on high-resistivity SOI substates, we observe a decrease with temperature similar to standard SOI substrates, however, a similar empirical exponential decay does not appear to fit well the data, suggesting that multiple mechanisms contribute to losses in high-resistivity SOI substrates.\\

We now focus on the frequency shifts vs temperature, as presented in Fig. \ref{fig:Temp}b. For bulk HR and trap-rich SOI substrates, there is no observable shift until the decrease associated to thermally induced quasiparticles, which starts to dominate above about 400 mK. We note that the decrease in $Q_i$ that we observe above about 250 mK in trap-rich SOI substrates is not associated to a measurable frequency shift. For standard SOI substrates, we observe a frequency shift above 200 mK, most likely linked to the associated decrease in $Q_i$, whose origin is unknown. Finally no clear trend is observed in high-resistivity SOI substrates.

\section{Discussion and simulations}

\begin{figure*}[t]
\centering
\includegraphics[width=0.5\textwidth]{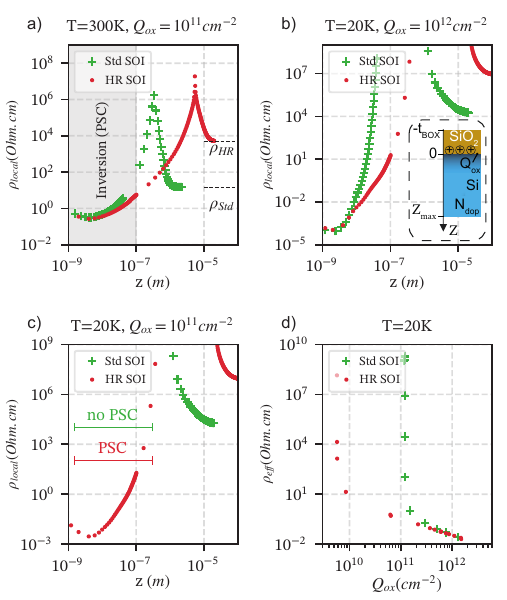}
\caption{\label{fig:simu} \textbf{Simulation  of parasitic surface conduction layer (PSC) in p-doped SOI substrates.} Calculation of the local charge density and resistivity in a structure composed of SiO$_2$ (thickness t$_{BOX}$=200~nm) and Si (thickness z$_{max}$=100~$\mu$m) (inset of b). Q$_{ox}$: positive charge defect density introduced at the interface. N$_{dop}$: dopant density in the Si. a) Local resistivity ($\rho_{local}$) as a function of the distance from the interface at room temperature for Q$_{ox}$=10$^{11}$cm$^{-2}$. For both standard SOI (Std SOI, N$_{dop}$=10$^{21}$ m$^{-3}$) and high-resistivity SOI (HR SOI, N$_{dop}$=3$\times$10$^{18}$ m$^{-3}$), there exists an inversion layer (PSC). 
Dotted lines: bulk resistivities ($\rho_{HR}$ and $\rho_{Std}$).
b) Local resistivity at 20~K for Q$_{ox}$=10$^{12}$cm$^{-2}$. Both substrates show a strong PSC but shallower in space for standard SOI. c)  Local resistivity at 20~K for Q$_{ox}$=10$^{11}$cm$^{-2}$. Here only the high-resistivity substrate shows a PSC, with the standard resistivity substrate being depleted all the way to the interface. d) Effective resistivity ($\rho_{eff}$) at 20~K calculated by integrating the local resistivity within a typical depth of 1~$\mu$m.}
\end{figure*}

\subsection{Summary of the experimental findings}

Our measurements yield multiple conclusions regarding microwave losses in SOI substrates. First and foremost, quality factors higher than 20 000 are demonstrated for superconducting resonators on trap-rich SOI substrates, which is 5 and 20 times better than standard and high-resistivity SOI substrates respectively, demonstrating the interest of trap-rich SOI substrates to mitigate microwave losses associated to PSC effects down to millikelvin temperatures. Losses in trap-rich SOI substrates should ultimately be dominated by dielectric losses associated to TLS in the buried silicon oxide layer. Using electromagnetic simulations, we estimated  that $Q_i=23~000$ is expected for $tan~\delta_{SiO_2} \simeq 3\times10^{-4}$, which is close to reported values for silicon oxide at 4K and down to millikelvin temperatures \cite{krupka2006measurements,o2008microwave}. This confirms that bulk $\textrm{SiO}_2$ losses is the dominant loss mechanism in trap-rich SOI substrates. Precise characterization of TLS losses through the $\beta$ value, associated to the efficiency of saturation of loss mechanisms, shows that absorption in $\textrm{SiO}_2$ is more difficult to saturate ($\beta \sim0.3$) compared to interface TLS which dominates in bulk silicon substrates ($\beta \sim 0.8$). Also, an additional mechanism results in an increase of losses in trap-rich SOI substrates for $T$ higher than 250 mK, but remains unclear. Investigation of the TLS switching dynamics versus temperature using $1/f$ permittivity noise spectroscopy could provide insights on the detailed mechanism involved in these variations \cite{ramanayaka2015evidence, burnett2014evidence,burin2015low}.\\

The significantly lower $Q_i$ in standard and high-resistivity SOI substrates compared to trap-rich SOI substrates points towards the existence of a PSC at millikelvin temperature, below the expected freeze-out of bulk charge carriers induced by dopants in silicon, typically around 50 K. This hypothesis is supported by the modeling of the substrates that we discuss in the following.
\subsection{Parasitic surface conduction simulations}
The calculations follow the approach described in refs.\cite{Rack2021a,Rack2021b}. We used a 1D Poisson Schrödinger solver to simulate the SOI structures with a 200~nm BOX and either a standard or high-resistivity substrate under the BOX (see the inset of Fig.~\ref{fig:simu}b). The Si substrates are p-doped  with a doping of 10$^{21}$ m$^{-3}$ (corresponding to a resistivity of  $\SI{15}{\ohm \cdot \centi\meter}$ at 300~K) to simulate standard resistivity SOI substrates, and a doping of 3$\times$10$^{18}$ m$^{-3}$ (corresponding to a resistivity of  $\SI{4800}{\ohm \cdot \centi\meter}$  at 300~K) to simulate  high-resistivity  SOI substrates. We solved the self-consistent Poisson Schrödinger equations for these materials at 300~K and 20~K. We accounted for the incomplete ionization of dopants at 20~K. We also considered a discrete dopant level \cite{Altermatt06}. Dopants can however be activated by the electric field at the interface between the BOX and the substrate. We thus computed the electron and hole carrier concentrations within the substrate, from the BOX interface (z=0 in  Fig.~\ref{fig:simu}) down to 100~µm in depth. We considered different oxide charge surface densities (Q\textsubscript{ox}) at the BOX/silicon interface. The corresponding local resistivity is finally obtained by multiplying the carrier density with the corresponding carrier mobility value, depending on temperature and doping level \cite{Morin54}.

In Fig.~\ref{fig:simu}a, we first present the room temperature case. We see that for a typical charge defect density of 10$^{11}$cm$^{-2}$ above the BOX/silicon interface, we observe a highly conducting inversion layer close to the interface for both high-resistivity and standard SOI substrates. This is the PSC. The spatial extension of the layer depends on the bulk doping level and increases when the doping decreases (high-resistivity SOI). Below the inversion layer, there exists a depletion layer with a very large local resistivity, before reaching the bulk silicon properties farther away from the interface. The situation gets more complex when lowering the temperature. In Fig.~\ref{fig:simu}b and c, we show simulations for two different defect sheet densities in both standard and high-resistivity SOI substrates. We see that at 20~K, i.e. below the dopant freeze-out temperature, the existence of the PSC depends very sensitively on both the bulk doping level and the charge defects density above the BOX/Si interface. There is a defect charge density threshold, which depends on the bulk doping level, below which the PSC does not exist. To summarize this, we show in Fig.~\ref{fig:simu}d, the estimated equivalent resistivity of the PSC as a function of the charge defect density. We observe a threshold at a defect density of 10$^{11}$cm$^{-2}$ (resp. 10$^{10}$cm$^{-2}$) for standard SOI substrate (resp. high-resistivity SOI substrates). In practice with a threshold as low as 10$^{10}$cm$^{-2}$, we can expect that there will always be an important effect of the PSC in high-resistivity SOI for realistic charge defect densities.   

\subsection{Discussion}
To go further in the analysis, we can attempt to match the measured $Q_i$ on standard and high-resistivity SOI substrates to electromagnetic simulations, including a typical 1~$\mu m$ thick conducting silicon layer below the BOX to model the PSC. From electromagnetic simulations, we obtained an effective resistivity of $\rho_{eff}$ $\sim$ 33~k$\Omega$.cm and $\sim$10~k$\Omega$.cm for standard and high-resistivity SOI substrates respectively. Comparing these values to Fig.~\ref{fig:simu}d, this would correspond to charge defect densities close to the threshold for the existence of the PSC. For high-resistivity SOI, it means a density around $10^{10}$cm$^{-2}$ which is the typical value expected in such substrate. For standard SOI, this would correspond to a larger defect density, around $10^{11}$cm$^{-2}$. 

We do not expect such large variation of the defect density between the two substrates. Thus, we tentatively conclude that in the case of standard resistivity SOI, other mechanisms dominate the microwave losses, possibly due to the doping of the silicon substrate. This is supported by the different temperature and power behaviors between the two types of substrates. 

In particular, we do not expect a strong temperature dependence for the PSC-associated losses. Losses in standard SOI are sensitive to temperature, with activation energy as low as 50~mK or less. The temperature dependence does not match a single activation energy ($E_a$), which would give a temperature variation following a $e^{\left( - \frac{E_a}{k_B T} \right)}$ dependence. Variable range hopping conduction also does not match the observed temperature dependence. The empirical exponential decay law we observe is not reported in literature, which suggests a significant contribution of disorder so that increasing the sample temperature amounts to probing a distribution of activation energies. Multiple sources of disorder can be at play here. Local disorder induced by the $\mathrm{Si/SiO_2}$ interface is expected to result in localized states whose energy lies close to the conduction band and whose signature can become relevant at cryogenic temperature \cite{ghibaudo1986transport, mott1987mobility}. In other words, shallow interface traps would be frozen at zero temperature but easily escaped from, resulting in decreasing effective trap density with increasing temperature \cite{galy2018cryogenic} and enhanced conduction. Such sub-100 mK activation energies were recently reported for dipoles in standard SOI substrates through their effect on spin qubits \cite{champain2025}. 

Another mechanism of bulk losses was recently identified in p-doped bulk silicon substrates \cite{zhang2024acceptor}. In that case, boron acceptors were shown to form TLS sites responsible for additional dissipation channels.
However, the $Q_i$ decay with temperature we observe does not match a TLS-like mechanism, and typical expected values of $Q_i$ associated to this mechanism would be more than one order of magnitude higher.\\

Having established that the losses in high-resistivity SOI substrates are limited by the PSC, we can now interpret the power and temperature dependence.  The variations of $Q_i$ are indeed peculiar in high-resistivity SOI substrates, as we observe clear bell shapes for both power and temperature dependence, unlike in standard SOI substrates. We propose an hypothesis explaining these differences, based on the behavior of the PSC at low temperature and in particular its spatial extension. In high-resistivity SOI substrates, the very low dopant density yields a PSC with a charge density profile extending far away from the interface (typically 1~$\mu$m). Carriers trapped to shallow defects could then be liberated by a temperature increase or microwave signal and contribute significantly (compared to the very low intrinsic dopant density) to the reduction of the spatial extension of the conductive PSC layer \cite{neve2012small}. This would provide an explanation for the bell shape we observed at low temperature in high-resistivity SOI substrates. 

In Table~\ref{tab:summary}, we summarize the discussion by indicating the dominant loss mechanisms for the studied SOI substrates. 
\\

\begin{table}[ht]
\begin{center}
\begin{tabular}{|p{6cm}|p{4cm}|p{4cm}|}

\cline{2-3}
\multicolumn{1}{c|}{} & T=300~K  & T=25~mK\\
\hline
Silicon on insulator with high-resistivity bulk& Parasitic surface conduction &Parasitic surface conduction \\
\hline
Silicon on insulator with standard resistivity bulk & Bulk resistivity and Parasitic surface conduction  &Bulk Si dissipation \\
\hline
Silicon on insulator with a trap-rich layer and high-resistivity bulk& Bulk resistivity & BOX dielectric losses \\
\hline

\end{tabular}
\end{center}
\caption{Dominant microwave dissipation mechanisms in SOI substrates.}
\label{tab:summary}
\end{table}

\section{Conclusion}
In this work, we investigate the mechanisms behind the microwave losses in different types of SOI substrates using superconducting circuits. We show that high-resistivity SOI substrates present large losses, due to the presence of the parasitic surface conduction layer, limiting internal quality factors to about $10^3$. Standard resistivity SOI substrates perform better but they are still limited by losses in the doped Si substrate. Trap-rich SOI substrates significantly outperform both since losses related to dopants in the substrate and the PSC are effectively suppressed. Internal quality factors above $2.10^4$ are reported in the single photon limit, only limited by losses in the SiO$_2$ layer. Improvements could be obtained by changing the participation ratio of the SiO$_2$ layer with a modification of the coplanar waveguide geometry dimensions. Nevertheless the quality factor reported here is already compatible with the use of these substrates for advanced quantum devices operating in cryogenic conditions, for instance electromechanical devices \cite{dieterle2016superconducting}  or microwave to optical converters \cite{zhao2025quantum}. These substrates will provide an enhanced flexibility due to the possibility to conserve the oxide underneath the active superconducting microwave part as it will improve both the mechanical integrity and heat dissipation \cite{kolvik2025,burger2026}.

\section*{Data Availability Statement}

The supporting data for this article are openly available from the Zenodo repository with the identifier \url{https://doi.org/10.5281/zenodo.21676679}.

\section*{Acknowledgments}
 We acknowledge the work of Julien Jarreau, Laurent Del-Rey and Didier Dufeu for the design and fabrication of the sample holders and other mechanical pieces used in the cryogenic system. We acknowledge discussions with Victor Champain, Xavier Jehl, Matias Urdampilleta and Simon Zihlmann. This work was supported by the French National Research Agency (ANR) in the framework of the STOUT project (ANR-21-CE47-0020). This work benefited from a French government grant managed by the ANR agency under the ‘France 2030 plan’, with reference ANR-23-PETQ-0005.
All SOI Wafers were provided by SOITEC, and HR bulk wafers by SILTRONIC within the framework of the MATQu project (ECSEL Joint Undertaking (JU) under grant agreement No 101007322.).

\bibliography{biblio}

\end{document}